\documentclass[aps,prl,onecolumn]{revtex4}
\usepackage{graphics}

\def \sds {\strut\displaystyle}

\def \v {{\bf v}}
\def \u {{\bf u}}
\def \x {{\bf x}}
\def \A {{\bf A}}
\def \K {{\bf K}}
\def \P {{\bf P}}

\def \tg {\tilde{g}}

\def \simless {\mathbin{\lower 3pt\hbox{$\rlap{\raise 5pt 
              \hbox{$\char'074$}}\mathchar"7218$}}} 
\def \simgreat {\mathbin{\lower 3pt\hbox{$\rlap{\raise 5pt 
              \hbox{$\char'076$}}\mathchar"7218$}}} 

\begin{document}
\title{3D Theory of the Plasma Cascade Instability\thanks{Work performed
under the auspices of the United States Department of Energy.}}

\author{M. Blaskiewicz\thanks{blaskiewicz@bnl.gov}\\
          BNL 911B, Upton, NY 11973, USA }

\begin{abstract}

  The plasma cascade instability (PCI) \cite{vl18,jma18,vl19,vlipac19,gangipac19,mmb19a,mmb19b} is a proposed mechanism for microbunching in electron beams without 
dipole magnets. Existing theory is limited to wave propagation that is orthogonal to the advective compression direction.
This work provides a theory allowing for wave propagation in arbitrary directions.
\end{abstract}
\maketitle
 
\section{introduction}
The plasma cascade instabilty (PCI)   \cite{vl18,jma18,vl19,vlipac19,gangipac19,mmb19a,mmb19b}   is a proposed mechanism for microbunching in electron beams without 
dipole magnets. If the theory bears out this process may well be very widespread, contributing to enhanced noise in a 
variety of systems employing electron beams. The actual system is quite complicated. This note considers what happens when an ion
is introduced into an infinite plasma that is undergoing expansion and contraction. 
 The first part of the paper considers the fluid limit and the Vlasov theory is presented in the second part. In both parts the original
partial differential equations are reduced to one dimensional equations which are straightforward to solve to high precision.
Actual solutions, which essentially require a full cooling design, are left to future work.

\section{fluid limit}
Consider a homogeneous, infinite, electron plasma.  It is described with Cartesian spatial coordinates $x_1,x_2,x_3$ and time $t$. 
Consider the Lorentz frame that moves with the beam and has a stable fixed point at the origin. 
The unperturbed plasma has a
velocity distribution
\begin{equation}
\v_0({\bf x},t) = \sum\limits_{j=1}^3 \hat{x}_j x_j \omega_j(t),
\label{aeq1}
\end{equation}
where $\hat{x}_j$ is a unit vector, and $\omega_j(t)$ has the units of frequency. As the $\omega_j$s vary in time the plasma expands and contracts in a similar way to the Hubble flow of galaxies.
The unperturbed density $n_0(\x,t)$ obeys
\begin{equation}
{\partial n_0 \over \sds \partial t} + \nabla \cdot ({\bf v}_0 n_0) = 0.
\label{aeq2}
\end{equation}
where  $\nabla$ is the gradient operator. Assume a spatially constant density $n_0 = n_0(t)$ which yields 
\begin{equation}
{d n_0 \over \sds d t} + (\omega_1(t)+\omega_2(t)+\omega_3(t)) n_0 = 0.
\label{aeq3}
\end{equation}
Defining $\omega_j(t) = \dot{\Phi}_j(t)$, where the dot denotes a time derivative,  gives $n_0(t) = \hat{n}_0 \exp(-\Phi_1(t) - \Phi_2(t)-\Phi_3(t))$
with $\hat{n}_0$ constant.
This defines our time dependent unperturbed distribution. The backround velocity distribution
is generated by a mixture of focusing from magnets, cavities and space charge forces. For the plasma cascade instability to manifest 
these variations need to be sufficiently robust. What this means in practice is left for future work. 
  For cooling systems with the beam propagating
along $x_3=z$ it is likely that $\omega_3\approx 0$ but it is kept to allow for general calculations. 

Now consider perturbations $\v_1(\x,t)$ and $n_1(\x,t)$. Work to first order in perturbation theory so the force and particle conservation equations are
\begin{eqnarray}
&&{\partial \v_1 \over \sds \partial t} + (\v_1 \cdot \nabla)\v_0 +  (\v_0 \cdot \nabla)\v_1 = q{\bf E}_1/m, \label{aeq4}\\
&&{\partial n_1 \over \sds \partial t} + \nabla \cdot ( \v_1 n_0 + \v_0 n_1) = 0, \label{aeq5}
\end{eqnarray}
where $q=-|e|$ is the electron charge, $m$ is the electron mass, and ${\bf E}_1$ is the electric field due to $n_1$
and the ion that we will consider cooling.
Next, solve Equations (\ref{aeq4}) and (\ref{aeq5}) using Fourier transforms with time dependent spatial wave numbers.
\begin{equation}
\v_1 = \tilde{\v}(t) \exp \left[ i \sum\limits_{j=1}^3 p_i \lambda_i(t) x_i \right]\equiv \tilde{\v}(t) \exp(i\Psi),
\label{aeq6}
\end{equation}
where the phase $\Psi(\x,t) = p_1\lambda_1(t)x_1 + \ldots$ is defined implicitly and the time dependent functions $\lambda_j(t)$ remain to be determined.
Since all the spatial dependence is in $\Psi$, take $n_1(\x,t) = \tilde{n}(t) \exp(i\Psi)$. 
Inserting these in (\ref{aeq4}) and (\ref{aeq5}) and defining ${\bf{E}}_1= \tilde{{\bf E}}(t)\exp(i\Psi)$ gives
\begin{equation}
{\partial \over\sds \partial t} \left[ \tilde{v}_j(t)e^{\sds i\Psi}\right]  + \tilde{v}_j \omega_j(t) e^{\sds i\Psi}
+ \left\{\sum\limits_{k=1}^3 x_k\omega_k {\partial \over\sds \partial x_k} \right\} 
\left[ \tilde{v}_j  e^{\sds i\Psi} \right] =  e^{\sds i\Psi}q\tilde{E}_j  /m, \label{aeq7}
\end{equation}
for $j=1,2,3$.   Particle conservation becomes
\begin{equation}
{\partial \over\sds \partial t} \left[ \tilde{n}_1(t)e^{\sds i\Psi}\right] + \tilde{n}_1(\omega_1+\omega_2+\omega_3) e^{\sds i\Psi}
+ \left\{  \sum\limits_{k=1}^3 x_k\omega_k {\partial \over\sds \partial x_k}   \right\} 
\left[ \tilde{n}_1  e^{\sds i\Psi} \right] + n_0(t)\left[ \sum\limits_{j=1}^3\tilde{v}_j {\partial \over\sds \partial x_j} \right]e^{\sds i\Psi}=0.
\label{aeq10}
\end{equation}
It is now clear how to choose the $\lambda_j$s. We demand
\begin{equation}
{\partial \over\sds \partial t}e^{\sds i\Psi} 
+ \left\{ \sum\limits_{k=1}^3 x_k\omega_k {\partial \over\sds \partial x_k} \right\} e^{\sds i\Psi}=0,
\label{aeq11}
\end{equation}
which leads to 
\begin{equation}
\dot{\lambda}_j + \omega_j \lambda_j=0,
\label{aeq12}
\end{equation}
where $j = 1,2,3$.   The solution is $\lambda_j(t) = \exp(-\Phi_j(t))$.  
Equations (\ref{aeq7}) through (\ref{aeq10}) become
\begin{eqnarray}
&& \dot{\tilde{v}}_j + \omega_j \tilde{v}_j = q\tilde{E}_j/m \label{ef1}\\
&& \dot{\tilde{n}}   + (\omega_1 + \omega_2 +\omega_3)\tilde{n} + n_0(t)\sum\limits_{k=1}^3 ip_k\lambda_k\tilde{v}_k =0\label{ef2}
\end{eqnarray}
To close the equations we use Gauss' law,  $\nabla \cdot {\bf E} = 4\pi q n$. Since everything varies as $\exp(i\Psi)\equiv \exp(i{\bf K}(t)\cdot{\bf r})$ we have
\begin{equation}
{\bf{E}} = -{ i {\bf{K}} \over \sds K^2} 4\pi q n + {\bf E}_{drive,{\bf K}},
\label{aeq13}
\end{equation}
where ${\bf E}_{drive,{\bf K}}$ is the spatial Fourier component of the electric field due to a driving ion. We note here that the motion
in the beam frame is assumed nonrelativistic. Otherwise there would be magnetic fields and retarded times, greatly complicating the problem. 
Additionally we assume that the ion can be treated in the impulse approximation with its position given by $\x = \x_0(t) = \x_0 + \v_0t$. 
 
Now
$$
{Q (\x -\x_0(t)) \over \sds |\x-\x_0(t)|^3} = -i \lambda_1\lambda_2 \lambda_3 {4 \pi Q \over \sds (2\pi)^3}\int d^3 p
{ (p_1\lambda_1,p_2 \lambda_2, p_3\lambda_3) \over\sds p_1^2 \lambda_1^2 + p_2^2 \lambda_2^2 +  p_3^2 \lambda_3^2 }
e^{\sds i \sum\limits_{j=1}^3 p_j\lambda_j (x_j - x_{0j}(t))},
$$              
which is easily checked by taking the divergence, using Gauss law on the left side and the definition of the 3 dimensional delta function on the right side. 

To keep our dynamics correct we need to sum quantities according to
$$ F_1(\x,t) = \int d^3 p \tilde{F}({\bf p},t) \exp(i p_1 \lambda_1 x_1 + \ldots)$$
where $F_1$ can be any of our small quantities. By doing this we keep all time dependent terms in $\tilde{F}\exp(i\Psi)$ which all obey Newton's Laws. 
With this convention the terms in equations (\ref{ef1}) through (\ref{ef2})
are related to the perturbed density and the external drive via
\begin{equation}
\tilde{\bf E} = -4\pi i { (p_1\lambda_1, p_2\lambda_2, p_3\lambda_3)\over\sds {\sum\limits_{j=1}^3   p_j^2 \lambda_j^2}}  \left\{  q\tilde{n}  
+ {H(t)Q\lambda_1\lambda_2\lambda_3  e^{\sds -i \sum\limits_{j=1}^3 p_j\lambda_j x_{0j}(t)} \over\sds (2\pi)^3}   \right\}.
\label{closer}
\end{equation}
Where $H(t)$ is the Heavyside function so that the ion is introduced at $t=0$. For cooling both $\tilde{v}$ and $\tilde{n}$ vanish at $t=0$ so Equations (\ref{ef1}) and (\ref{ef2}) 
with the definition (\ref{closer})
 are ordinary differential equations with simple boundary conditions. Next, the tricks learned here will be used in the Vlasov analysis.

\section{Vlasov limit}

For the Vlasov analysis we take Equation(\ref{aeq1}) for the average velocity as before.
This velocity is driven by the acceleration
\begin{equation}
\A_0(\x,t) = \sum\limits_{j=1}^3 \hat{x}_j x_j k_j(t) 
\label{eq2}
\end{equation}
We wish to find a distribution with these parameters. We will take an unperturbed
distribution of the form $f_0(\x,\v,t) = f_0(H(\x,\v,t))$ with
\begin{equation}
H = \sum\limits_{j=1}^3 { \alpha_j(t) \over\sds 2}\left(v_j-\omega_jx_j\right)^2.
\label{eq3}
\end{equation}
 While one could choose to make $f_0$ a function of each of the three terms in the sum of Equation (\ref{eq3}), the simpler approach will be taken here. 
The Vlasov equation is
\begin{equation}
{\partial f_0 \over \sds \partial t} + \v \cdot {\partial f_0 \over \sds \partial \x} + \A_0(\x,t)\cdot  {\partial f_0 \over \sds \partial \v} =0,
\label{eq4}
\end{equation}
and since $f_0 = f_0(H)$, $H$ satisfies Equation (\ref{eq4}) as well. Insert $H$ for $f_0$ in Equation  (\ref{eq4}). The resulting terms proportional
to $x_j$ and $v_j$ are
\begin{equation}
-\alpha_j(v_j-\omega_jx_j)\dot{\omega}_j x_j + \dot{\alpha}_j(v_j-\omega_jx_j)^2/2 - v_j\alpha_j\omega_j(v_j-\omega_j x_j)+ x_jk_j\alpha_j(v_j-\omega_jx_j)=0.
\label{eq4a}
\end{equation}
Setting the coefficients of $x_j^2$, $x_jv_j$ and $v_j^2$ to zero we find Equation (\ref{eq4}) is satisfied if
\begin{equation}
\dot{\alpha}_j = 2\alpha_j\omega_j,\hspace*{1cm}\dot{\omega}_j+\omega_j^2 = k_j
\label{eq5}
\end{equation}
for $j=1,2,3$. If Equations (\ref{eq5}) are satisfied then any  function $f_0(H)$ will satisfy Equation (\ref{eq4}). For physical solutions we
require $f_0(H)d^3xd^3v$ to be the number of electrons in the phase space volume $d^3xd^3v$. 

We will use first order perturbation theory with $f = f_0 + f_1$ so that
\begin{equation}
{\partial f_1 \over \sds \partial t} + \v \cdot {\partial f_1 \over \sds \partial \x} + \A_0(\x,t)\cdot  {\partial f_1 \over \sds \partial \v}
+  \A_1(\x,t)\cdot  {\partial f_0 \over \sds \partial \v}  =0,
\label{eq6}
\end{equation}
where $\A_1$ is the acceleration created by $f_1$ and by the ion seeding the instability. Using the results of the previous section we introduce time
dependent spatial wave numbers and assume a perturbation where the spatial density of the electrons varies as 
\begin{equation} 
n_\P(\x,t) = \hat{n}_\P(t)\exp\left( i\sum\limits_{j=1}^3 P_j\lambda_j(t) x_j \right)\equiv  \hat{n}_\P(t)\exp(i\Psi(\x,t))
\label{eq7}
\end{equation}
where $\dot{\lambda}_j + \omega_j\lambda_j=0$ and the $P_j,s$ are constant in time. The total perturbed density from all wavenumbers is
$$n_1(\x,t)=\int  n_\P(\x,t)d^3P.$$ 

Consider an ion implanted at $t=0$ and located at $\x = \x_0(t)=\x_0 + \v_0t$. The acceleration it generates is given by (cgs units)
\begin{eqnarray}
&& \A_{I}(\x,t) = {qQ\over\sds m} {\x-\x_{0}(t) \over\sds |\x-\x_0(t)|^3} \nonumber \\
&& = - i  {4\pi qQ\over\sds m (2\pi)^3} \int d^3P
\lambda_1\lambda_2\lambda_3  {(P_1\lambda_1,P_2\lambda_2,P_3\lambda_3)  \over\sds \sum_j P_j^2 \lambda_j^2} \exp\left( i\sum_m P_m \lambda_m (x_m - x_{0m}(t))\right),
\label{eq8}
\end{eqnarray}
where $q=-|e|$ is the electron charge, $m$ is its mass, and $Q$ is the charge on the ion. The net acceleration for wavenumber $\P$ due to both the ions and electrons is 
$\A_{\P}(x,t)= \tilde{\A}(t)\exp(i\Psi(\x,t))$ with
 \begin{equation}
\tilde{\A}(t) = -4\pi i{ {(P_1\lambda_1,P_2\lambda_2,P_3\lambda_3)} \over\sds {\sum_m P_m^2\lambda_m^2}}
\left\{ {q^2 \over\sds m }  \hat{n}_\P(t)    + {qQ\over\sds (2\pi)^3m} \lambda_1\lambda_2\lambda_3  \exp\left( -i\sum_m P_m \lambda_m x_{0m}(t)\right)       \right\}.
\label{eq9}
\end{equation}

To solve the Vlasov equation we consider a single $\P$.  Consider the Ansantz 
$$f_\P(\x,\v,t) = {df_0 \over\sds dH} g(v_1-\omega_1x_1,v_2-\omega_2x_2,v_3-\omega_3x_3,t)\exp(i\Psi(\x,t)).$$
Notice that the $x_j$ dependence in $g$ and $f_0$ only shows up as $v_j-\omega_jx_j$ so it drops out after integrating over $v_j$. This generates the correct spatial dependence
for $n_\P(\x,t)$. For convenient notation define $u_j = v_j-\omega_j x_j$ and remember that
$${\partial g(\x,\v,t) \over \sds \partial t} =  {\partial g(\u,t) \over \sds \partial t} +  
{\partial g(\u,t) \over \sds \partial \u}\cdot  {\partial \u(\x,\v,t) \over \sds \partial t}.$$

Plugging into the Vlasov eq one finds 
\begin{equation}
{\partial g(\u,t) \over \sds \partial t}+\sum\limits_{j=1}^3 igP_j\lambda_j u_j - \omega_j u_j {\partial g \over \sds \partial u_j} + \alpha_ju_j \tilde{A}_j =0
\label{eq10}
\end{equation}
To proceed we multiply the last term on the right of Equation (\ref{eq10}) by $\delta(t-t_0)$ with the intention of integrating over $t_0$ later.
\begin{equation}
{\partial \tilde{g}(\u,t,t_0) \over \sds \partial t}+\sum\limits_{j=1}^3 i\tg P_j\lambda_j u_j - \omega_j u_j {\partial \tg \over \sds \partial u_j} + \alpha_ju_j \tilde{A}_j\delta(t-t_0) =0
\label{eq10a}
\end{equation}

We look for solutions of the form
$$\tilde{g}(\u,t,t_0)=H(t-t_0){\bf q}(t)\cdot\u \exp(i\K(t)\cdot\u)$$
where $H(t-t_0)$ is 1 for $t \ge t_0$ and zero otherwise.  Inserting this expression into (\ref{eq10}) and using the same sort of tricks used to solve (\ref{eq4a})
 we find that (\ref{eq10a})  is satisfied if
% array with odes and boundary conditions.
\begin{eqnarray}
&&\dot{K}_j + P_j\lambda_j(t)- \omega_j(t)K_j=0,\hspace*{5mm}{\rm with}\hspace*{2mm} K_j(t_0)=0 \label{eq11a}\\
&&\dot{q}_j - \omega_j(t)q_j=0,\hspace*{15mm}{\rm with}\hspace*{2mm} q_j(t_0)=-\alpha_j(t_0) \tilde{A}_j(t_0). \label{eq11b}
\end{eqnarray}
To bring the pieces together define the general solution $M_j(t)$ as the solution to Equation (\ref{eq11a}) but with the boundary condition $M_j(0)=0$.
Also define the phases $\Phi_j(t)$ so that $\omega_j(t) = \dot{\Phi}_j(t)$. With these definitions $K_m(t,t_0) =  M_m(t)-M_m(t_0)\exp[ \Phi_m(t)-\Phi_m(t_0)]$ and  
$q_j(t,t_0)=-\alpha_j(t_0) \tilde{A}_j(t_0)\exp [ \Phi_j(t)-\Phi_j(t_0)]$. This yields
\begin{equation}
g(\u,t)= \sum\limits_{j=1}^3 \int\limits_0^{t} dt_0 q_j(t,t_0)u_j \exp\left( i \sum\limits_{m=1}^3 u_m K_m(t,t_0)\right)
\label{eq13b}
\end{equation}
To close the equations we note that the only unknown in $\tilde{A}_j(t)$ is 
\begin{eqnarray}
&&\hat{n}_{\P}(t) = \int d^3 u   {\partial f_0 \over \sds \partial H}(\u,t)g(\u,t) \label{eq14a}\\
&&= -\sum\limits_{j=1}^3 \int\limits_0^{t}dt_0 \alpha_j(t_0)\tilde{A}_j(t_0)\exp\left(  \Phi_j(t)-\Phi_j(t_0)\right) 
\int d^3u  f_0'\left(   \sum\limits_{m=1}^3 \alpha_m(t) u_m^2 /2\right)u_j \exp\left( i \sum\limits_{k=1}^3 u_k K_k(t,t_0)\right),
\label{eq14b} \\
&&= -\sum\limits_{j=1}^3 \int\limits_0^{t}dt_0 \tilde{A}_j(t_0) G_j(t,t0),
\label{eq14c} \\
&&= \int\limits_0^{t}dt_0 \left[ {q^2\over\sds m}\hat{n}_{\P}(t_0)+D_I(t_0)\right] \sum\limits_{j=1}^3 \left( {4\pi i P_j \lambda_j(t_0) \over\sds \sum\limits_{m=1}^3 \lambda_m^2(t_0)P_m^2}G_j(t,t_0)\right)
\label{eq14d}
\end{eqnarray}
where
\begin{eqnarray}
&&G_j(t,t_0) = \alpha_j(t_0)\exp\left(  \Phi_j(t)-\Phi_j(t_0)\right) 
\int d^3u  f_0'\left(   \sum\limits_{m=1}^3 \alpha_m(t) u_m^2 /2\right)u_j \exp\left( i \sum\limits_{k=1}^3 u_k K_k(t,t_0)\right),
\label{eq15a}\\
&& = -{i\alpha_j(t_0)K_j(t,t_0)\over\sds \alpha_j(t)}\exp\left(  \Phi_j(t)-\Phi_j(t_0)\right) 
\int d^3u  f_0\left(   \sum\limits_{m=1}^3 \alpha_m(t) u_m^2/2 \right)\exp\left( i \sum\limits_{k=1}^3 u_k K_k(t,t_0)\right).
\label{eq15b}
\end{eqnarray}
and
\begin{equation}
D_I(t)= {qQ\over\sds(2\pi)^3m} \lambda_1(t)\lambda_2(t)\lambda_3(t)  \exp\left( -i\sum_m P_m \lambda_m(t) x_{0m}(t)\right).
\label{eq16}
\end{equation}
We now have a Volterra equation of the second kind for $\hat{n}_{\P}$. While exact solutions look hopeless a numerical solution should be straightforward.
We note that changing the integration variable in Equation (\ref{eq15b}) to $z_i = \sqrt{\alpha_i}u_i$ turns it into the Fourier transform of a spherically symmetric function
for which a wide range of exact solutions are available.
\section{Connection to previous work}
If we set $\omega_i=0$ then these results should reduce to those in~\cite{gang}.  To show this we set
$$d_i(t) = {Z \over\sds(2\pi)^3} \lambda_1(t)\lambda_2(t)\lambda_3(t)  \exp\left( -i\sum_m P_m \lambda_m(t) x_{0m}(t)\right),$$
$$g({\bf K}(t,t_0)) = \int d^3u  f_0\left(   \sum\limits_{m=1}^3 \alpha_m(t) u_m^2/2 \right)\exp\left( i \sum\limits_{k=1}^3 u_k K_k(t,t_0)\right),$$
$$R(t,t_0) = {1\over \sds  \sum\limits_{m=1}^3 \lambda_m^2(t_0)P_m^2} \sum\limits_{j=1}^3 { P_j \lambda_j(t_0) \alpha_j(t_0)K_j(t,t_0)\over\sds \alpha_j(t)}\exp\left(  \Phi_j(t)-\Phi_j(t_0)\right),$$
where $Z$ is the atomic number of the ion.
Now we have
\begin{equation}
\hat{n}_{\P}(t) = { 4\pi q^2 \over\sds m} \int\limits_0^{t}dt_0 \left[ \hat{n}_{\P}(t_0)-d_i(t_0)\right] R(t,t_0)g({\bf K}(t,t_0)).
\label{final}
\end{equation}
 When $\omega_i=0$, $R(t,t_0) = t_0-t$, $\lambda_i=1$, and ${ \bf K}(t,t_0) = (t_0-t){\bf P}$. Make these substitutions, 
account for a difference in Fourier transform conventions, and include the fact that there is a constant velocity offset between reference frames. One finds that Equation~(\ref{final}) here is
equivalent to Equation (8) in \cite{gang}.
\section{Logarithmic Concerns}
If one were to actually solve Equation (\ref{final}) and integrate it over all $\P$ a logarithmic divergence for large $\P$ would result. 
This is due to the fact that first order perturbation theory overestimates the momentum transfer for small impact parameters. It is possible to correct for this
singularity by making the substitution
$$d_i(t)\rightarrow d_i(t)\exp\left( -b_{min}^2 \sum\limits_{j=1}^3 \lambda_j^2(t)P_j^2 \right), $$
where $b_{min}$ is the minimum impact parameter and the exponential form is just a suggestion. Problems associated with large impact parameters
will not occur since the charge will be Debye shielded.

\end{document}